\documentclass[conference]{IEEEtran}

\usepackage{amssymb,amsfonts}
\usepackage{graphicx}
\usepackage{listings}
\usepackage{hyperref}
\usepackage{float}
\restylefloat{table}
\usepackage{flushend}

\usepackage{xcolor}
\usepackage[normalem]{ulem} 			
\usepackage{ifthen}
\usepackage{xspace}

\newboolean{showcomments}
\setboolean{showcomments}{true}

\ifthenelse{\boolean{showcomments}}
{
	\newcommand{\nb}[3]{
		{\colorbox{#2}{\bfseries\sffamily\scriptsize\textcolor{white}{#1}}}
		{\textcolor{#2}{\textsf\small$\blacktriangleright$\textit{#3}$\blacktriangleleft$}}}
	 
	\newcommand{\bnote}[2]{\fbox{\color{blue}\bfseries\sffamily\scriptsize#1}
    	{\color{blue}\textsf\small$\blacktriangleright$\textit{#2}$\blacktriangleleft$}}
	\newcommand{\old}[1]{{\color{gray}\sout{#1}}} 
	\newcommand{\del}[1]{\old{#1}} 
	\newcommand{\ins}[1]{{\textcolor{blue}{\uline{#1}}}} 
	\newcommand{\ugh}[1]{{\textcolor{red}{\uwave{#1}}}} 
	\newcommand{\chg}[2]{{\textcolor{red}{\sout{#1}}}{\ra}\textcolor{blue}{\uline{#2}}} 
	
	\newcommand{\fix}[1]{\bnote{FIX}{#1}}
}{
	\newcommand{\bnote}[2]{}
	\newcommand{\nb}[3]{}
	\newcommand{\old}[1]{}
	\newcommand{\del}[1]{}
	\newcommand{\ins}[1]{}
	\newcommand{\ugh}[1]{}
	\newcommand{\chg}[2]{}
	
	\newcommand{\fix}[1]{}
}

\newcommand{\hide}[1]{}

\graphicspath{{images/}}

\newcommand{\eg}[1]{\textit{e.g.}\xspace}
\newcommand{\ie}[1]{\textit{i.e.}\xspace}

\usepackage{url}
\makeatletter
\def\url@leostyle{%
  \@ifundefined{selectfont}{\def\UrlFont{\textsf}}{\def\UrlFont{\small\sffamily}}}
\makeatother
\makeatother

\definecolor{verylightgray}{rgb}{0.95, 0.95, 0.95}

\newcommand{\code}[1]{\colorbox{verylightgray}{\texttt{#1}}}

\newcommand{\esope}{Esope\xspace}
\newcommand{\oldf}{Fortran-77\xspace}
\newcommand{\newf}{Fortran-2003\xspace}

\newcommand{\linebreakand}{%
\end{@IEEEauthorhalign}
\hfill\mbox{}\par
\mbox{}\hfill\begin{@IEEEauthorhalign}
}

\begin{document}

\title{Parsing \oldf with proprietary extensions}

\author{\IEEEauthorblockN{Younoussa Sow}
\IEEEauthorblockA{\textit{DTIPD \textit{Framatome}} \\
Paris, France \\
younoussa.sow@framatome.com}
\and
\IEEEauthorblockN{Larisa Safina}
\IEEEauthorblockA{\textit{Univ. Lille, Inria, CNRS, Centrale Lille,} \\
\textit{UMR 9189 CRIStAL}\\
F-59000 Lille, France \\
larisa.safina@inria.fr}
\and
\IEEEauthorblockN{Léandre Brault}
\IEEEauthorblockA{\textit{DTIPD \textit{Framatome}} \\
Paris, France \\
leandre.brault@framatome.com}
%
\and
\IEEEauthorblockN{Papa Ibou Diouf}
\IEEEauthorblockA{\textit{DTIPD \textit{Framatome}} \\
Paris, France \\
papa-ibou.diouf@framatome.com}
\and
\IEEEauthorblockN{St\'ephane Ducasse}
\IEEEauthorblockA{\textit{Univ. Lille, Inria, CNRS, Centrale Lille,} \\
\textit{UMR 9189 CRIStAL}\\
F-59000 Lille, France \\
stephane.ducasse@inria.fr}
\and
\IEEEauthorblockN{Nicolas Anquetil}
\IEEEauthorblockA{\textit{Univ. Lille, Inria, CNRS, Centrale Lille,} \\
\textit{UMR 9189 CRIStAL}\\
F-59000 Lille, France \\
nicolas.anquetil@inria.fr}
}

\maketitle
\IEEEpeerreviewmaketitle

\begin{abstract}
Far from the latest innovations in software development, many organizations still rely on old code written in ``obsolete'' programming languages.
Because this source code is old and proven it often contributes significantly to the continuing success of these organizations.
Yet to keep the applications relevant and running in an evolving environment, they sometimes need to be updated or migrated to new languages or new platforms.
One difficulty of working with these ``veteran languages'' is being able to parse the source code to build a representation of it.
Parsing can also allow modern software development tools and IDEs to offer better support to these veteran languages.
We initiated a project between our group and the Framatome company to help migrate old \oldf with proprietary extensions (called \esope) into more modern Fortran.
In this paper, we explain how we parsed the \esope language with a combination of island grammar and regular parser to build an abstract syntax tree of the code.
\end{abstract}


\section{Introduction}

A significant amount of the software used nowadays in critical sectors of the industry (\eg\ nuclear, engineering, or banking sectors) is written in languages that are considered to be ``veterans'' in computer science,  such as Fortran, COBOL, Ada, etc.
Nowadays, Cobol remains dominating in the financial domain being used in 43\% of US banking systems and 95\% of ATMs, and estimated as having around 220 billion lines of code\footnote{\url{http://fingfx.thomsonreuters.com/gfx/rngs/USA-BANKS-COBOL/010040KH18J/index.html}}.
Fortran prevails as the number one language for scientific, engineering, and high-performance computing (HPC) domains due to its performance, numerical computation capabilities, and extensive legacy code base.
For example, even the modern SciPy Python library is actually a wrap-up around old Fortran (and C) libraries.

But working with veteran languages raises specific problems:
\begin{itemize}
\item They tend to have limited support for modern software engineering practices (testing, code-commenting, documentation, etc.) and do not benefit from such tools as code quality and security evaluation, code completion, code auto-generation, or refactoring.

\item They have small abstraction power having been designed at a time when computers were much smaller and slower and developers needed fine control over the compiled code to optimize it (\eg\ it was normal in Fortran to take memory allocation alignment into account when declaring variables\footnote{\url{https://fftw.org/doc/Allocating-aligned-memory-in-Fortran.html}}).
They may also lack dynamic data allocation, forcing developers to produce complex and entangled solutions.

\item Veteran languages are not taught anymore, thus a recent report\footnote{\url{https://permalink.lanl.gov/object/tr?what=info:lanl-repo/lareport/LA-UR-23-23992}} identified as a real threat the difficulty to find top-rate Fortran computer scientists in the future;

\item Hardware platforms running this code are becoming more and more outdated. The code is not always ported on new platforms and does not profit from the new technologies with better performance.
\end{itemize}

Yet, organizations are forced to support irreplaceable old code-base, if only to adapt it to new needs or security standards.
This prompted the creation of different extensions of these languages, for example for Fortran: Parametric Fortran, Fortran-S, Vienna Fortran, \esope.
Because they have a more restricted distribution, these extensions present the same risks listed above to an even higher degree.
This creates a strong demand for methods and tools for parsing, modeling, or migrating such veteran languages and/or their extensions.

We are experiencing these difficulties as we were called on a project to migrate \esope/\oldf projects to \newf by the Framatome company\footnote{\url{https://www.framatome.com/en/}}.
Framatome designs, builds, and services nuclear steam supply systems.
The first roadblock of this project is to be able to parse the \esope/\oldf source code (\esope being a proprietary extension of the Fortran language).

In this paper we explain what are the challenges linked to parsing veterans languages in general, and \oldf in particular.
In our case, these challenges are made harder by the fact that we are dealing with an extension of Fortran that has only one known parser available.
We propose an approach that allowed us to parse this language at a low cost.

The paper is structured as follows:
In Section~\ref{sec:background} we present the background of the project and the specificities of the two languages we have to deal with: \esope and \oldf.
In Section~\ref{sec:related-work} we list the existing work on Fortran migration and some parsing technologies such as Island Grammars.
In Section~\ref{sec:model-esope}, we explain why parsing these veteran languages is difficult, we list our constraints for a Fortran parser, and review the existing solutions.
In Section~\ref{sec:solution} we propose a solution to parse the \esope extension language without the need to go into the trouble of creating a whole parser for the language.
We present the result of evaluating our solution on a small but representative and independent project in Section~\ref{sec:evaluation}.
Finally, we close the paper with a discussion of future work (Section~\ref{sec:future-work}) and the conclusion (Section~\ref{sec:conclusion}).

\section{Background}
\label{sec:background}

\subsection{Fortran}

For historical reasons, Fortran is one of the main programming languages for scientific computing, so much, that the SciPy Python library is actually a wrap-up around old Fortran code.

There are several dialects of the language, standardized at different times: early versions (Fortran-66, \oldf, Fortran-90, Fortran-95), and modern versions, adding for example concepts from Object-Oriented programming (\newf, Fortran-2008, Fortran-2018).
There are also non-standard extensions:
\begin{itemize}
\item Parametric Fortran \cite{Erwi07a} enables the creation of Fortran extensions using parameter structures, which can be referenced in Fortran programs to indicate the dependence of program sections on these parameters.
\item Fortran-S \cite{Bodi93a} enhanced \oldf with directives to specify parallelism (shared data structures and parallel loops).
\item Fortran-D \cite{Fox90a} is an extension to Fortran90 enhanced with data decomposition specifications also used for writing parallel programs, that was later integrated to a newer version, Fortran 2008.
\item Vienna Fortran 90 \cite{Benk92a} is a language extension of Fortran 90 which enables the user to write programs for distributed memory multiprocessors using global data references only.
\item \esope is a Fortran extension created by the CEA\footnote{Alternative Energies and Atomic Energy Commission or CEA (Commissariat à l'énergie atomique et aux énergies alternatives) \url{https://www.cea.fr}}
to allow (i) complex data structures use, (ii) automatic memory management, and (iii) automatic swap management.
\end{itemize}

Having been the standard for 13 years, \oldf is historically important and has a very large code base.
This is the standard on which \esope was based and that is much used at Framatome.

Fortran programs come in two forms: fixed-column (early versions) and free-form (since Fortran-90) formats.
The free-form formats resemble current programming practices where the position of a character on the line has no formal significance and can be used for example for statement indentation.

The fixed-column inherits characteristics from the old punched card systems that were used to enter programs in a computer:
\begin{itemize}
\item Lines are limited to 72 characters (or columns);
\item A ``C'' or ``*'' as the first character of the line (column 1) marks a comment line (the line is ignored by the compiler);
\item Otherwise, the first 5 characters of a line (columns 1 to 5) are for a label field: a sequence of digits that can be referred to by GOTO statements;
\item Column 6 is for a continuation field: a character other than a blank or a zero there causes the line to be taken as a continuation of the preceding line.
\item Columns 7 to 72 serve as the statement field;
\item Anything from column 73 is ignored and can be used for comment.
\end{itemize}

Many scientific or engineering programs today still use the fixed-column format because they were written a long time ago or by people who had learned to program in early versions of Fortran.
We will see that it has consequences on migration projects.

\subsection{\esope}

Before Fortran 90, there were some issues in data management:
\begin{itemize}
\item No user-defined data structures (other than arrays and matrices);
\item No dynamic allocation of memory;
\item No automatic memory swap (memory paging).
\end{itemize}

At the end of the seventies, the CEA created an extension to Fortran 77.
The goal was to create a programming language allowing (1) Complex data structures, (2) Automatic memory management, (3) Automatic swap management.
The objective of \esope is to facilitate the management of data and to allow the notion of an object by the structuring of data like struct in C.
The motivation was to have a set of data within a single variable. 
This notion was unknown to \oldf.
This is how an entity called SEGMENT was added, as well as primitives for manipulating it.

A segment is:
\begin{itemize}
\item A group of Fortran variables defined by the programmer;
\item Referenced by a single variable called POINTER.
Knowing the pointer is enough to access all the variables contained in the structure.
\end{itemize}

\esope is an extension to Fortran meaning that before learning how \esope works, one should know the basics of Fortran, especially control structures, subroutine calls, COMMON statements, etc.
The functionality of \esope allows one to define, initialize, copy, change, and suppress a segment, get and change a segment's dimension and declare multiple pointers attached to a segment.

We give in Listing~\ref{lst:esope-example} an example of \esope code for the management of books and users inside a library.
\esope introduces:
\begin{itemize}

\item A new type of definition structure, the ``segment''.
The example in Listing~\ref{lst:esope-example} declares a structure for a \code{user} with a \code{name} (line 5) and an array of borrowed books (\code{ubb}, line 6).
The size of the array is dynamically defined by the variable \texttt{ubbcnt}.
The ``field'' definitions have the same syntax as regular variable definitions in Fortran;

\item A new data type, \code{pointer} for a pointer variable (\texttt{ur}) pointing to a \texttt{user} structure (line 8);

\item New statements to manipulate the segments (\eg\ an ``instance'' of \code{user} is created at line 11 with \code{segini}).
There are six new statements: \emph{segini}, \emph{segact}, \emph{segadj}, \emph{segdes}, \emph{segprt}, and \emph{segsup};

\item New library functions (\eg\ \code{actstr}, \code{ajpnt}, or \code{mypnt}, not shown in the listing);

\item New notations to access the members of a segment (dot notation, line 12) and to get the size of an array (slash notation, line 13, gets the size of the first dimension of array \code{ur.ubb}).
\end{itemize}

We will explain in our solution how we deal with each of these extensions that generate potential problems for parsing.

\begin{lstlisting}[language=Fortran, label=lst:esope-example, caption=Example of \esope code (see text for explanations)]
      SUBROUTINE NEWUSER(LIB,NAME)
      IMPLICIT NONE
      INTEGER UBBCNT
      SEGMENT, USER
       CHARACTER*40 UNAME
       INTEGER UBB(UBBCNT)
      END SEGMENT
      POINTEUR UR.USER
C the user does not have a book yet 
      UBBCNT = 0
      SEGINI, UR
      UR.UNAME = NAME
      WRITE(*,*) UR.UBB(/1)
[...]
\end{lstlisting}

\subsection{Aging tools}

The \esope transpiler takes an \esope source code (\esope new instructions inter-mixed with standard \oldf source code) and generates pure \oldf code with memory manipulation instructions.
For this, it makes strong assumptions on how the \oldf compiler (Intel Fortran compiler) allocates memory.
The latest versions of the compiler are not able to optimize the source code generated, which is a huge drawback for Fortran applications.
It is feared that the future version of the compiler will not be able to handle the code at all.

In the prevision of this, a research project was launched to offer an automatic translation solution from \esope sources to modern Fortran (\newf) allowing an object-oriented approach.
A preliminary study showed that it is possible to perform this translation from \esope sources to \newf manually. 
However, the large number of sources to be translated and their size do not allow manual translation to be considered, which explains the need for an automatic approach.
The objective is therefore to offer solutions for the automatic translation of codes from the \esope overlay into modern Fortran. 

\section{Related Work}
\label{sec:related-work}

Maintaining and providing an up-to-date infrastructure for proprietary programming languages, domain-specific programming languages, or language dialects remains a real problem in reverse engineering~\cite{Jong01a,Deurs98a,SNEED2019162}.

Jonge and Monajemi~\cite{Jong01a} suggest several options for resolving maintenance problems.
They include outsourcing the development in the given (veteran) language, simplifying the development process, and migrating to a similar standardized programming language.
They do mention that the last one (our case) can be very challenging.
According to the same article and~\cite{Lamm01a},  the main effort in building a parser for migrating a language is in (re)defining a grammar, as opposed to building tools around it (building parse trees, pretty printing, etc.).

Evolving a DSL requires adapting the compiler of a hosting language which in turn requires corresponding skills and background from engineers, and time and money investments from companies.
Lammel and Verhoef in~\cite{Lamm01a} state that implementing a high-quality Cobol parser can take up to three years, and adapting an existing parser to deal with language changes/dialects can take three to five months.

Island grammars are a possible solution to the definition of a new grammar/parser.
They are a partial definition of a language's grammar, a common way to identify and annotate statements of an embedded language ~\cite{Moon01a,Moon02a}.
An island grammar is characterized by its production rules, which provide comprehensive descriptions of specific constructs known as ``islands,'' alongside more lenient rules that encompass the remaining elements referred to as ``water.''
Island grammars work where normal grammars do not (\eg\ in handling incomplete and syntactically incorrect code, embedded code written in other programming language or dialect, code using preprocessors, etc.) and help to avoid the tedious and expensive writing of a complete language grammar and parser.

\subsection{Parsing and Migration of Fortran}

There exist numerous projects dedicated to parsing and converting code written in Fortran to other programming languages including Python~\cite{Bysi16a}, JVM Bytecode or Java~\cite{Fox97a,Seym01a,Dool99a}, C/C++~\cite{Feld95a,Feldm91,Gros12a, f2cpp}, Pascal~\cite{Frea81a}, Basic~\cite{Cari92a}, Ada~\cite{Slap83a,Pars88a}, Algol~\cite{Prud77a} and others.

Bysiek et al~\cite{Bysi16a} present a semi-automatic transpilation for a subset of Fortran 77/90/95 to Python 3 to retain Fortran-level performance. The paper, however, does not give any details on the parsing process in transpilation.

The first automatic Fortran to C conversion, a program called f2c, was presented by Feldman \emph{et al.} in~\cite{Feld95a,Feldm91}.
The creation of the tool was motivated by the need to run Fortran program on a machine that had a C compiler but no Fortran compiler.
It allowed to better express certain functionalities that were easier to do in C than in Fortran (\eg\ storage management, some character operations, arrays of functions, heterogeneous data structures, and calls that depend on the operating system), to profit from the C tools like linters and verifiers for consistency and portability checks etc.\footnote{Here, we would like to underline the fact that problems of old programming languages which have motivated this work, still persist, thirty years after creating the f2c tool.}.
The tool is based on Feldman's previous f77 compiler~\cite{Feld79a} that originally parsed Fortran to an intermediate representation, but was tweaked to produce a C parse tree used as an input for the second pass of the portable C compiler.
We review f2c parsing capability in Section~\ref{sec:review-compilers}.

In~\cite{Gros12a} Grosse-Kunstleve et al. show the problems of integrating Fortran code to modern Object-Oriented Programming environments (\eg\ extensive use of global variables, communicating using intermediate data files etc.) and present a Fortran to C++ source-to-source conversion tool FABLE.
Comparing to~\cite{Feld95a} generated C++ code had better readability and the tool itself could be integrated in modern modular systems.

For Java, there exist two solutions, both called f2j and focused on porting of Fortran libraries to profit from Java's portability and proliferation. The first one, by Fox et al~\cite{Fox97a} is based on the Fortran to C conversion~\cite{Zhan97a,zhen94a} and parses an input Fortran program to an AST, converts Fortran AST to C AST, and then generates java source code from it. The second one, by Dongarra, Seymour, and Doolin~\cite{Seym01a,Dool99a} focuses on parsing Fortran to a JVM-bytecode to facilitate translation of Fortran GOTO statements (that do not always have a direct translation to a corresponding Java code) and to explore possible code optimizations on the byte-code level. The tool does lexing/parsing to build a complete AST and to use it further for optimization, type assignments, and code generation purposes.

\section{Modeling Fortran/\esope source code}
\label{sec:model-esope}

Our migration project will use a model-driven approach, a proven technique for source code modification.
The challenge of the project addressed in this paper is how to build a model of the Fortran/\esope source code.

\subsection{Challenge: Building a model of the code}

We elected to work with the Moose platform \cite{anqu20a} that offers tools for software analysis and manipulation.
It has a specific AST (Abstract Syntax Tree) meta-model independent of the programming language.
However, Moose does not have a Fortran importer.
This is a bottleneck of the model-driven approaches that a model of the system must be built, which involves parsing the source code.
Creating custom parsers for most programming languages is a non-trivial endeavor that can be even more difficult for veteran languages \cite{Lamm01a}.

\esope source is currently handled by a pre-processor written in \esope itself.
Although it could seem the most natural way to go, we choose not to use this pre-processor because it would have meant learning two new languages (\esope and \oldf) that lag behind modern programming techniques and especially do not facilitate handling abstract concepts and data structures.

Another conceivable solution would have been to apply the \esope pre-processor to obtain pure \oldf source code and parse it to find back the \esope instructions that led to the generated Fortran code.
However, the  pre-processor generates very low-level, memory manipulation, code that does not allow to recover the high-level instructions.
For example, Listing~\ref{lst:esope-preproc} shows the preprocessed result of the \esope code of Listing~\ref{lst:esope-example} (note that Fortran is not case sensitive).
Lines 4 to 16 in Listing~\ref{lst:esope-preproc} are the translation of the segment definition and pointer declaration (lines 4 to 8 in Listing~\ref{lst:esope-example}).
Lines 18 to 23 in Listing~\ref{lst:esope-preproc} are the translation of the \code{segini} instruction (line 11 in Listing~\ref{lst:esope-example}).
Lines 24 to 26 (Listing~\ref{lst:esope-preproc}) are the translation of the assignment (line 12 in Listing~\ref{lst:esope-example}).
Note that line 26 (Listing~\ref{lst:esope-preproc}) has a character in column 6, indicating it is the continuation of the preceding line.
The character itself (a ampersand) is not part of the continuation, it is only a marker.
Lines 27 and 28 (Listing~\ref{lst:esope-preproc}) are the translation of the assignment (line 13 in Listing~\ref{lst:esope-example}).

\begin{lstlisting}[language=Fortran, label=lst:esope-preproc, caption=\oldf result of the preprocessed \esope code from Listing~\ref{lst:esope-example}]
      SUBROUTINE NEWUSER(LIB,NAME)
      IMPLICIT NONE
      INTEGER UBBCNT
C      segment, user
C      POINTEUR UR.USER
c the user does not have a book yet
      COMMON/OOOCOM/OOT,OOV(2),OO_001,OO_002,OO_003,
      OO_004
      INTEGER*8OOW(1)
      INTEGEROOV,OOO,OO1,OO2,OO3,OO4,USER,OO5,UR
      INTEGEROOI(1)
      INTEGER*8OOT
      CHARACTER*4OOH(1)
      EQUIVALENCE(OOV(1),OOW(1),OOI(1),OOH(1))
      INTEGEROO_001(2),OO_002(2),OO_003(2)
      CHARACTER*4OO_004(2)
      UBBCNT = 0
C      SEGINI, UR
      CALLOOOWIN(OO4,0,'NEWUSE 10 UR ',OO1,13+UBBCNT)
      OO_001(-0002+OOW(OOT+OO1)+1)=40
      OO_002(-0004+OOW(OOT+OO1)+2)=13
      OO_003(-0006+OOW(OOT+OO1)+3)=UBBCNT
      UR=OO1
C      UR.UNAME = NAME
      OO_004(-0008+OOW(OOT+UR)+1)(OOV(2)+12+1:OOV(2)+12
     &+OOV(OOW(OOT+UR)+1))=NAME
C      WRITE(*,*) ur.ubb(/1)
      WRITE(*,*) OO_005(-0010+OOW(OOT+UR)+5)
[...]
\end{lstlisting}

We therefore looked for a \oldf parser, open-source, that we could adapt to accept \esope source code (\ie \oldf with \esope extensions).

Note that the first versions of Fortran (the early fifties, work lead by J.W. Backus\footnote{Also known for the Backus-Naur notation}) slightly pre-dates the theory of formal languages (Chomksy, 1954) and as such did not have a formal grammar to describe the language.
For example, Fortran compilers are expected to accept programs where some ``tokens'' do not need to be separated by white spaces which makes parsers cumbersome to write.
There are several cases of this in Listing~\ref{lst:esope-preproc}, for example line 11, ``\code{INTEGEROOI(1)}'' for ``\code{INTEGER OOI(1)}'', or line 19, ``\code{CALLOOOWIN(...)}'' for ``\code{CALL OOOWIN(...)}''.
This is one of the difficulties of Fortran parsing that had to be taken into account.

\subsection{Criteria for an \esope/Fortran parser}
\label{sec:criteria}

A good candidate able of transforming both \esope code and \oldf code in the form of an AST must offer the following:
\begin{itemize}
\item Allow to easily convert its AST representation to Moose AST format.
An acceptable solution for this is to output the parser AST in a standard data exchange format (eg: JSON or XML) and load it in Moose;

\item Keep the comments of the analyzed source code.
Migration between languages implies not only technical challenges of preserving the same functionality, performance, or security of the resulting code but also of preserving the original ``knowledge'' of the development team~\cite{Brag23a,Verh19a}.
Our goal must be to preserve the structure, identifiers, and comments of the current code to the greatest extent possible;

\item Give access to the positions in the source file of the AST nodes as several tools of Moose use this information;

\item Be open-source to be adaptable to the needs of the project.
\end{itemize}

We analyzed several tools that are able to parse Fortran and/or \esope source code and produce an intermediate representation in the form of AST: compilers, dedicated parsers, static analyzers, and code converters. We summarize our results in the next section.

\subsection{Fortran compilers}
\label{sec:review-compilers}

\textbf{gfortran\footnote{\url{https://gcc.gnu.org/wiki/GFortran}}} is a free Fortran compiler included in the GNU Compiler Collection (GCC) suite.
It fully implements the Fortran95/2003/2008/2018 standards and was supporting \oldf until recently (the support ceased in gcc-4.0).
It can parse \oldf generated by \esope.
After the parsing phase, gfortran can output an AST, however, in a ``not standard'' format which therefore requires additional processing and creating an additional grammar/parser for it.
It does not give access to the comments and positions of the instructions in the source file.

\textbf{ifort\footnote{\url{https://www.intel.com/content/www/us/en/developer/tools/oneapi/fortran-compiler.html\#gs.0p7v9b}}} is a proprietary compiler developed by Intel.
It was included in the study as it is used by the Framatome engineers to compile their sources.
However, there is no easy way to obtain an AST as well as other information we require.

\textbf{lfortran\footnote{\url{https://lfortran.org}}} is an open-source Fortran compiler that is constructed on the foundation of LLVM.
lfortran offers AST in a clear readable form that can also be exported to JSON.
Unfortunately, it has full support only for Fortran2018 and cannot be used for this project\footnote{After this survey, we discovered that the latest version of lfortran (version 0.19.0) should be able to parse Fortran-77
(\url{https://lfortran.org/blog/2023/05/lfortran-breakthrough-now-building-legacy-and-modern-minpack/}).
We come back to this in Section~\ref{sec:future-work}}.

\textbf{flang\footnote{\url{https://github.com/flang-compiler/flang}}} also referred to as ``Classic Flang,'' is a Fortran compiler intended for use with LLVM.
It represents an open-source iteration of pgfortran, which is a commercial Fortran compiler produced by PGI/NVIDIA.
It produces AST but in a format requiring additional work to be imported.

\textbf{F2c~\cite{Feld95a}} is a free \oldf compiler and \oldf to C code translator.
Developed in the 90s, it became a common mean to compile Fortran code and is used in many migration tools or compilers like gfortran.
Before performing the translation between languages, f2c generates a parse tree, however, we did not find a way to export it.
We considered making a direct translation of the \oldf/\esope code to Fortran2003 code without having to go through an AST (the process is described in~\cite{Feld95a}) but the changes we will have to do look too large to be handled without a model of the code.
There existed a tool to improve the readability of the source code obtained by f2c, which is, unfortunately, no longer available.

\subsection{Fortran-specific parsers}

\textbf{fortran-src\footnote{\url{https://github.com/camfort/fortran-src}}}~\cite{Orch13a} is a tool written in Haskell that provides lexing, parsing, and basic analysis for Fortran66/77/90/95/2003 as a part of a bigger CamFort (Cambridge Fortran infrastructure) project for reengineering and verification of scientific Fortran programs.
Fortran-src outputs an AST expressed in Haskell algebraic data types which required parsing to obtain an easy-to-use format.
However, Camfort recently provided JSON output of their AST, which makes it a good candidate for our project.
Fortran-src also allows keeping comments and positions in source files.

\textbf{Open Fortran Parser\footnote{\url{https://github.com/OpenFortranProject/open-fortran-parser}}} (OFP) consists of a parser and associated tools for Fortran2008 based on the ANTLR (ANother Tool for Language Recognition) grammar for Fortran.
It outputs AST in XML format, however, it is not suited to use with \oldf/90 and its output did not seem reliable for our purposes.

\textbf{ANTLR\footnote{\url{https://www.antlr.org}}}(ANother Tool for Language Recognition)~\cite{Parr07a} is a generator of front-ends for compilers which, starting from a grammar in Backus–Naur form (BNF) produces lexical and syntactic analyzers.
There exist community-written grammars for different versions of Fortran, but not for the code generated by \esope.
Indeed, \esope does not impose to separate tokens in the code which ends up causing the lexical analyzer to fail (\eg\ ``INTEGERVAR'' was not parsed into ``INTEGER'' and ``VAR'').
The solution would therefore be to find a way to modify expressions in a grammar used by a lexical analyzer which for the moment has not been conclusive.

\textbf{BNFC\footnote{\url{https://hackage.haskell.org/package/BNFC}}} (or the BNF Converter) is an open-source tool used in compiler construction to generate the front end of a compiler based on a labeled BNF grammar.
Initially designed for generating Haskell code, it can also generate other languages (\eg\ Ada, C/C++, Java, etc).
Based on the provided grammar, the tool produces several components, including an AST, and generates lexer and parser generator files that can be used with ANTLR and other tools.
However, our attempts to compile the Fortran source code generated by \esope were not successful.

\textbf{ROSE\footnote{\url{http://rosecompiler.org/ROSE_HTML_Reference/index.html}}}~\cite{Quin11a} is a compiler infrastructure that is open-source, designed for creating tools that perform program transformation and analysis on a source-to-source basis.
It is specifically developed for large-scale applications written in Fortran 77/95/2003, C, C++, OpenMP, and UPC.
Unlike most other research compilers, ROSE aims to allow non-experts to take advantage of compiler technologies to build their own custom software analyzers and optimizers.
At the time of this writing, we did not yet evaluate ROSE as a solution for our project.
It seems a serious candidate.

\subsection{Other parsers}

\textbf{PetitParser2\footnote{\url{https://kursjan.github.io/petitparser2/}}}~\cite{Kurs13a}
PetitParser2 is an open-source framework for building parsers developed by Jan Kurš. It is based on a unique combination of four alternative parsing methodologies:
\begin{enumerate}
    \item scannerless parsers;
    \item parser combinators;
    \item parsing expression grammars;
    \item Packrat analyzers.
\end{enumerate}
PetitParser2 allows one to define island grammars and would be capable of recognizing \esope instructions contained in an \esope source file. 

\textbf{ESOPE} is the tool used by Framatome that allows translation from \esope programming language to \oldf.
It is a preprocessor that is written in \esope itself.
It generates \oldf source code (as illustrated in Listing~\ref{lst:esope-preproc}).

All the above-listed solutions have been analyzed according to our stated criteria.
The results of this analysis are synthetized in Table~\ref{tbl:recap}.
The column titles in it correspond to:
\begin{enumerate}
    \item \textbf{F77}: Idiomatic \oldf support;
    \item \textbf{ESP}: Support for \oldf generated by \esope;
    \item \textbf{AST}: Offers an AST as output;
    \item \textbf{CMT}: Keeps comments;
    \item \textbf{POS}: Gives access to the positions in the nodes file of the AST;
    \item \textbf{OS}: Open-source tool.
\end{enumerate}

\begin{table}[h!]
    \centering
    \begin{tabular}{l@{ }c@{ }c@{ }c@{ }c@{ }c@{ }c}
        \hline
        Parseurs & F77 & ESP & AST & CMT & POS & OS\\
        \hline
        gfortran       & X & X & X &   &   & X        \\
        ifort          & X & X &   &   &   &        \\
        lfortran       & X &   & X & X & X & X       \\
        flang          &   &   &   &   &   &          \\
        f2c            & X &   &   &   &   &        \\
        Fortran-src    & X & X & X & X & X & X       \\
        OFP            &   &   & X &   &   & X       \\
        ANTLR          & X & X & X &   &   &              \\
        BNFC           & X & X & X &   &   & X       \\
        ESOPE          & X & X &   &   &   &        \\
        PetitParser2   &   & X & X & X & X & X       \\
        \hline
    \end{tabular}
    \caption{Tools under analysis}\label{tbl:recap}
\end{table}

Given the criteria mentioned and the tests carried out on all the sources provided, we distinguish two promising tools for parsing Fortran code: gfortran and fortran-src as they are capable of analyzing \oldf.
We chose the latter due to its ability to keep both the comments and the positions of the instructions as well as due to having AST presented in easily-parsable JSON format.

\section{Proposed Solution}
\label{sec:solution}

Since finding an \esope parser fulfilling our needs is difficult, we chose to follow another path.
We propose to work in three steps:
\begin{enumerate}
\item ``de-Esopify'' the \esope/Fortran code by creating pure \oldf source with \esope ``annotations'' (in comments);
\item parse annotated \oldf code to get its AST (with the comments);
\item process the generated AST to recover the \esope constructs/instructions.
\end{enumerate}

\subsection{``De-Esopify'' the \esope/Fortran code}

To remove \esope specific constructs and instructions, we defined an island grammar \cite{Moon02a} parser.
This parser rewrites the \esope source code into ``annotated'' \oldf code.
For this, it recognizes the \esope specific constructs and instructions (island grammar), and converts them either to Fortran comment or to valid Fortran code:
\begin{itemize}
\item Definition of a \code{segment}:
  \begin{itemize}
  \item The first line of the segment (line 4 in Listing~\ref{lst:esope-example}) is commented out so that we will be able to recover it later.
  We use a special marker ``\code{c@\_}'' that is unlikely to be found in normal programs.
  Remember that a ``c'' in the first column indicates a comment line in Fortran;
  \item the definition of the segment members (``attributes'', lines 5 and 6 in Listing~\ref{lst:esope-example}) are kept since they are valid variable definitions in Fortran;
  \item the end line of the segment (line 7) is also commented out.
  It would be syntactically valid (``END'' can be followed by anything), but would close something that was not opened in the Fortran code.
  \end{itemize}
  
\item The pointer definition is not valid Fortran as the \code{pointer} type is not known, therefore, the line is commented out

\item The six new \esope statements (\emph{segini} --on line 11--, \emph{segact}, \emph{segadj}, \emph{segdes}, \emph{segprt}, and \emph{segsup}) are again commented out;

\item The new \esope functions (\eg\ \code{actstr}, \code{ajpnt}, or \code{mypnt} are kept because they are already valid Fortran functions.
They will have to be treated when migrating the code, but they don't raise any parsing problems;

\item The dot notation to access segment ``attributes'' is replaced by a special expression: \code{ur.uname} becomes \code{D\_\_(ur,uname)} (``D'' for dot).
\code{D\_\_} is a valid Fortran identifier and unlikely to be found in real code.
The generated expression can have two interpretations in Fortran, which we will discuss in the next section;

\item Simlarly, the ``slash notation'' (\code{array(/n)}) to get the size of \code{array}'s \emph{n}th dimension (line 13 of Listing~\ref{lst:esope-example}) is replaced by another special expression: \code{ur.uname(/1)} becomes \code{S\_\_(D\_\_(ur,uname),1)} (``S'' for ``slash'').
The result is a syntactically correct Fortran expression (with the same caveat as above).
\end{itemize}

The result of ``De-Esopifying'' the code example of Listing~\ref{lst:esope-example} is given in Listing~\ref{lst:de-esopify}.
Note that this code is syntactically correct for \oldf, but semantically incorrect.
This is not a problem as we only wish to get the AST from the code to be able to analyze it.

\begin{lstlisting}[language=Fortran, label=lst:de-esopify, caption=``De-Esopification'' of the code example of Listing~\ref{lst:esope-example}. The\\ result is syntactically valid Fortran code.]
      subroutine newuser(lib,name)
      implicit none
      integer ubbcnt
c@_  segment, user
       character*40 uname
       integer ubb(ubbcnt)
c@_  end segment
c@_  pointeur ur.user
c the user does not have a book yet 
      ubbcnt = 0
c@_  segini, ur
      D__(ur,uname) = name
      bor = S__(D__(ur,ubb),1)
[...]
\end{lstlisting}

\subsection{Parsing the annotated Fortran code}

There is no difficulty here once we have a \oldf parser that matches our restrictions (see Section~\ref{sec:criteria}).
The annotated Fortran code is valid and should result in a valid Fortran AST.
We chose a parser that keeps comments in the AST and the annotations will be kept this way.

\subsection{Recover the \esope constructs/instructions}

Once we have the AST of the pure Fortran code, we transform it into an \esope/Fortran AST.
For this, we navigate the AST, looking for our annotations (line comment nodes).
We recognize our annotations by the special comment marker that we created (``\code{c@\_}'').
Each case is handled appropriately:
\begin{itemize}
\item A ``\code{c@\_   segment...}'' comment marks the start of a segment definition.
Its sibling nodes in the AST should be variable declaration statements and a ``\code{c@\_   end segment}'' comment.
We remove all these nodes (the two comments and the declaration statements in-between) from their parent and create a new special node \code{EsopeSegmentDefinitionNode} and give it the variable declaration statements as children.
 
\item A ``\code{c@\_   pointeur...}'' comment marks the declaration of pointer variable.
This node is replaced by an \code{EsopePointerVariableDeclaration} for the variable(s).
This involves a trivial parsing of the comment line where the \code{pointeur} keyword is followed by a word, a dot, and another word.
The first word is the name of the variable and the second is the segment type pointed to.
Optionally, several pointers may be declared on the same line (separated by commas), but this does not make the parsing much more difficult.

Note that a segment definition (previous item) may contain itself a pointer ``attribute''.
In such a case, the two annotations are treated successively to return the correct result (a \code{EsopeSegmentDefinitionNode} containing a \code{EsopePointerVariableDeclaration}).

\item A ``\code{c@\_   segini...}'' comment marks a special \esope instruction.
It is replaced by a new node \code{EsopeSegmentIntruction} that takes the name of the \esope instruction (\emph{segini}, \emph{segact}, \emph{segadj}, \emph{segdes}, \emph{segprt}, and \emph{segsup}) and, as a child, the pointer variable(s) treated.
Again this involves some trivial parsing of the comment line where the \esope command is followed by a comma-separated list of variable names.

\item The new \esope functions (\eg\ \code{actstr}, \code{ajpnt}, or \code{mypnt}) are not modified for now, but we could check all function call nodes to see whether they invoke a special \esope function or not.

\item The new dot and slash notations of \esope are found in the AST in two forms:
  \begin{itemize}
  \item When the ``\code{D\_\_}'' or ``\code{S\_\_}'' is an expression (right-hand side of an assignment for example) it will be seen in the AST as a call to a ``\code{D\_\_}'' or ``\code{S\_\_}'' functions.
  So ``\code{var = D\_\_(ptr,attribute)}'' will be seen in the AST as an assignment with a variable \code{var} on the left-hand side and a function call ``\code{D\_\_}'' with two arguments (\code{ptr} and \code{attribute}) on the right-hand side.
  In this case, we replace the function call node by a node \code{EsopeAttributeAccess} with two children (the pointer and the attribute).
 
  \item When the ``\code{D\_\_}'' or ``\code{S\_\_}''  is a variable (left-hand side of an assignment), it will be seen in the AST as a ``statement function'', which is a concise way in Fortran to define a function performing only one computation and returning its value.
  So ``\code{D\_\_(ptr,attribute) = value}'' will be seen in the AST a statement function where the function \code{D\_\_} has two parameters (\code{ptr} and \code{attribute}) and a body which is the \code{value}.
  In this case, we replace the statement function node by a node \code{EsopeAttributeAccess} with two children (the pointer and the attribute).
  \end{itemize}

\end{itemize}

With these transformations of our special annotations in the original \oldf AST, we obtain a new AST containing \esope specific nodes and representing accurately the initial \esope source code.

\section{Evaluation}
\label{sec:evaluation}

We evaluated our solution on the BookLibrary application, a toy \esope program.
The application was created independently by Framatome developers long before this project.
It contains 10 \esope source files.
Its interest lies not in the size of the source code (around 500 LOC in total) but in the fact that it uses all the different \esope-specific instructions and constructs.
Note that the running example of this paper (Listing~\ref{lst:esope-example} and following) comes from this application.

Table~\ref{tab:bookstore} lists some descriptive statistics about the BookLibrary.
For the 10 files of the example, it gives the number of lines of code (column \emph{LOC}), the number of segment definitions (column \emph{segment}), number of pointer definitions (column \emph{pointer}), number of uses of one of the six \esope instructions (segini,\ldots, in column \emph{instr.}), number of uses of the ``dot notation'' (column \emph{dot}), and number of uses of the ``slash notation'' (column \emph{slash}).
The three segment definitions of the example (book, user, and library) appear in the ``struc.inc''.
This file is included in all ``.E'' files that manipulate ``instances'' of the segments.
 
\begin{table}[htbp]
\begin{center}
\caption{Descriptive statistics on the files of our validation}
\label{tab:bookstore}
\begin{tabular}{lcc@{ }c@{ }c@{ }c@{ }c}
\hline
file & LOC & segment & pointer & instr. & dot & slash \\
\hline
struc.inc & 31 & 3 & 0 &  0 &  0 & 0 \\[1ex]
borbk.E   & 65 & 0 & 3 &  5 &  5 & 3 \\
findbk.E  & 54 & 0 & 3 &  4 &  4 & 2 \\
findur.E  & 55 & 0 & 3 &  4 &  5 & 3 \\
libpnt.E  & 73 & 0 & 4 & 10 & 16 & 3 \\
main.E    & 35 & 0 & 1 &  0 &  0 & 0 \\
newbook.E & 52 & 0 & 4 &  6 &  7 & 2 \\
newlib.E  & 39 & 0 & 3 &  3 &  0 & 0 \\
newuser.E & 53 & 0 & 5 &  6 &  4 & 2 \\
relbk.E   & 80 & 0 & 4 &  5 &  7 & 3 \\
\hline
\end{tabular}
\end{center}
\end{table}

\begin{figure}[htbp]
\begin{center}
\includegraphics[width=\linewidth]{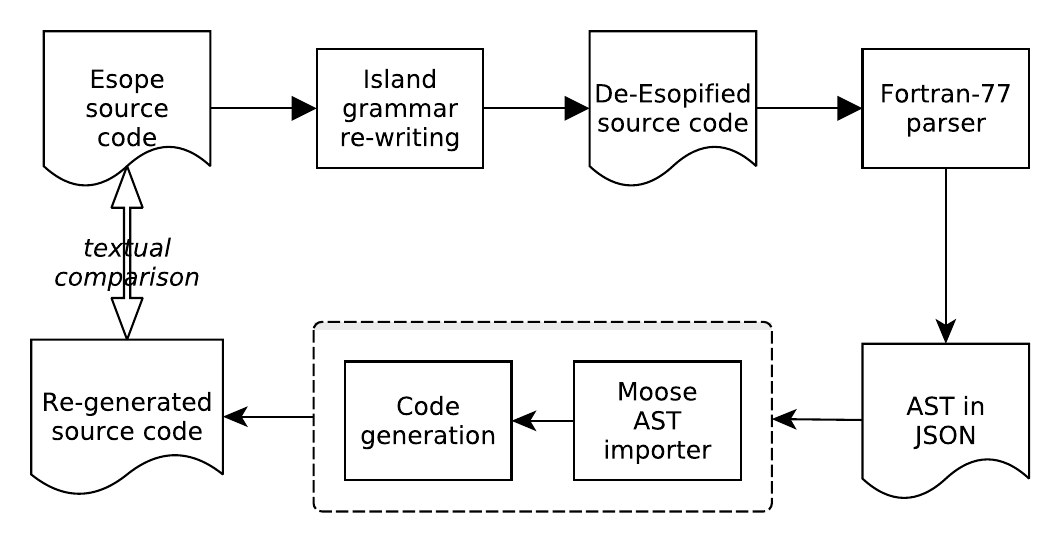} 
\end{center}
\caption{Round-trip validation of our chain of tools to parse \esope/\oldf source code and rebuild it from the AST}
\label{fig:roundtrip}
\end{figure}

To validate our model of the \esope/Fortran source code, we use a ``loose round-trip'' comparison (see Figure~\ref{fig:roundtrip}).
The round-trip comparison consists in taking a source code, modeling it as an AST in Moose, then regenerate the source code from this AST and compare it with the original code.
If the model is correct and complete, the generated source code will be the same as the initial source code.

There are several problems linked to this validation:
\begin{itemize}
\item Whitespaces and empty lines in source code are often not significant (apart from the six spaces at the beginning of the lines in Fortran), for example ``\code{( a .lt. 5 )}'' is equivalent to ``\code{(a.lt.5)}'' despite the differences in whitespaces.
Similarly, empty lines are not significant from the compiler's point of view, but they are from source code comparison.

\item Also Fortran is not case-sensitive, so that ``\code{if (a .lt. 5) then}'' is equivalent to ``\code{IF (A .lt. 5) THEN}''.

\item Fortran is a somewhat flexible language offering several different syntaxes to express the same semantics.
Comments can be expressed by a ``C'', ``c'', or ``*'' in the first column of a line, or a continuation line can be marked by any character in column 6.

\item Still for flexibility, in Fortran, a subroutine declaration with no parameter can be written with or without parentheses: ``\code{subroutine insgrp()}'' can also be written ``\code{subroutine insgrp}''.
\end{itemize}

A perfect regeneration of code should regenerate exactly the whitespace and empty lines that were in the original code although they rarely are part of the AST.

In our AST, we don't store the actual tokens found in the code because this is not relevant to our goal.
For example, we don't store the actual ``IF'' token found in the code, but the fact that there is a  ``FortranIfStatement''.

The rationale is that even if for the migration, we need to generate source code with the same structure (functions, instructions, flow of control) of the original one, and the same documentation (comments, identifiers) so that the developers will be able to recognize their source code and understand it.
But the \esope constructs and specificities will disappear, so obviously some differences are expected.
We will format the code in a way that makes it easily readable (source code pretty-printer or beautifier) and according to developers' wishes, but we will not aim to reproduce \emph{exactly} the same format (which might actually not be correctly formatted in the first place).

Therefore, in the AST, we only indicate there is a comment line or a continuation line without keeping the exact character that marked it, or we just store a list of parameters that can be empty.
When regenerating source code, we put a ``C'' in column 1 for comments, a ``\&'' in column 6 for continuation lines, we write ``IF'' (not ``if''), and ``()'' for an empty list of parameters.
If this is not how the original code was formatted, a strict round-trip comparison will signal these differences.

For the evaluation, we, therefore, applied some modifications to the original source code:
\begin{itemize}
\item Apply a preprocessor to deal with the ``\code{\#include}''  instruction;

\item Ignore whitespace, some tools like the GNU diff program have an option for that (\url{https://www.gnu.org/software/diffutils/manual/html_node/White-Space.html}, \code{--ignore-all-space});

\item Convert both texts to compare to lowercase;

\item Remove all ``\texttt{()}'' of subroutine declarations (note: it is easier to remove it from both texts than adding it to the original code when it is missing);

\item Replace ``*'' at the beginning of a line (first column) with a ``C''.
\end{itemize}
This does not solve all problems, but greatly reduces the number of differences so that the remaining can be very quickly checked manually.

We applied our chain of tools (see Figure~\ref{fig:roundtrip}) to all the files of the BookLibrary example and the re-generated files were all equivalent to the original ones.

As an additional verification, we executed a small ``test case'' that uses the BookLibrary (creates books and users, loan and return books) and prints the state of the library after each action.
We compared the printed information of this small test on the original and re-generated code and both outputs are strictly equal.

\section{Future work and discussions}
\label{sec:future-work}

The markers for annotations (\code{c@\_}), dot and slash notations (\code{D\_\_(...)}, and \code{S\_\_(...)}) were chosen to minimize the probability of finding them in real source code.
We are currently just assuming they do not appear in the company's code.
For more security, we should add a test on the source code before applying the ``de-Esopification''.
This is a trivial task as a simple text search in the source code will tell us if these strings are found.
In such a case, a warning should be raised and different markers should be chosen.

Because the Fortran parser that we are currently using (camfort) is permissive, we did not check for the length of the generated lines.
When a line contains several cases of dot and slash notations, there is a high risk that the ``de-Esopified'' line will exceed 72 characters (we replace the single dot character by six characters: ``\code{a.b}'' becomes ``\code{D\_\_(a,b)}'').
With more strict Fortran parsers, this should be checked and a continuation line (a line with a character in column six) should be introduced.
We have so far resisted doing it because we wish to keep the ``de-Esopified'' code as close as possible to the \esope code.

Since our initial analysis of Fortran parsers (see Section~\ref{sec:model-esope}), we discovered that lfortran evolved to handle \oldf, that ROSE looked like a good candidate, and that there was another parser (SYNTAX\footnote{\url{https://fr.wikipedia.org/wiki/SYNTAX}}) that could also be used.
This is not a threat to our results since our approach accepts any Fortran parser.
Changing the parser only requires adapting the Moose AST importer (see again Figure~\ref{fig:roundtrip}) which is not a difficult step.

\section{Conclusions}
\label{sec:conclusion}

In industry, a vast amount of companies are still using software written in old programming languages that often do not conform to modern standards and best practices and are not handled by modern development environments and tools.
However, companies are forced to keep such software since there does not exist a reliable solution to migrate it.
The situation gets worse if it concerns the rarely used programming languages: dialects, proprietary languages, DSL, etc.
In this paper, we present a part of the project made for the Framatome company in migration of the code written in such language called \esope which is an extension of Fortran-77.
Here, we focus on parsing \esope: underlying challenges and possible solutions.
We present an analysis of existing parsers for \esope/Fortran-77 and propose a solution consisting of: 
(1) using an island-grammar parsing phase that remove \esope constructs from the source code and creating pure Fortran-77 source with commented \esope ``annotations'',
(2) parsing annotated (``de-esopified'') Fortran-77 code to get its AST (with the comments), and
(3) processing the generated AST to recover the \esope constructs.
We evaluate our solution on the 500 LOC example provided by Framatome, that is, however, complete as it covers all possible \esope constructs.

The work described in this paper is only the very first step of the project that should lead to the migration of \esope code into \newf code

\bibliographystyle{IEEEtran}
\bibliography{rmod,others,local}

\begin{thebibliography}{10}
\providecommand{\url}[1]{#1}
\csname url@samestyle\endcsname
\providecommand{\newblock}{\relax}
\providecommand{\bibinfo}[2]{#2}
\providecommand{\BIBentrySTDinterwordspacing}{\spaceskip=0pt\relax}
\providecommand{\BIBentryALTinterwordstretchfactor}{4}
\providecommand{\BIBentryALTinterwordspacing}{\spaceskip=\fontdimen2\font plus
\BIBentryALTinterwordstretchfactor\fontdimen3\font minus
  \fontdimen4\font\relax}
\providecommand{\BIBforeignlanguage}[2]{{%
\expandafter\ifx\csname l@#1\endcsname\relax
\typeout{** WARNING: IEEEtran.bst: No hyphenation pattern has been}%
\typeout{** loaded for the language `#1'. Using the pattern for}%
\typeout{** the default language instead.}%
\else
\language=\csname l@#1\endcsname
\fi
#2}}
\providecommand{\BIBdecl}{\relax}
\BIBdecl

\bibitem{Erwi07a}
Z.~F. Martin~Erwig and B.~Pflaum, ``Parametric fortran: Program generation in
  scientific computing,'' \emph{Journal of Software Maintenance and Evolution},
  vol.~19, no.~3, pp. 155--182, 2007.

\bibitem{Bodi93a}
\BIBentryALTinterwordspacing
F.~Bodin, L.~Kervella, and T.~Priol, ``Fortran-s: A fortran interface for
  shared virtual memory architectures,'' in \emph{Proceedings of the 1993
  ACM/IEEE Conference on Supercomputing}, ser. Supercomputing '93.\hskip 1em
  plus 0.5em minus 0.4em\relax New York, NY, USA: Association for Computing
  Machinery, 1993, pp. 274--283. [Online]. Available:
  \url{https://doi.org/10.1145/169627.169732}
\BIBentrySTDinterwordspacing

\bibitem{Fox90a}
G.~C. Fox, S.~Hiranandani, K.~Kennedy, C.~Koelbel, U.~Kremer, C.-W. Tseng, and
  M.~L.~C. Wu, ``Fortran d language specification,'' Dept. of Computer Science,
  Rice University, Tech. Rep. CRPC\#TR90–141, 1990.

\bibitem{Benk92a}
S.~Benkner, B.~Chapman, and H.~Zima, ``Vienna fortran 90,'' in
  \emph{Proceedings Scalable High Performance Computing Conference SHPCC-92},
  1992, pp. 51--59.

\bibitem{Jong01a}
M.~de~Jonge and R.~Monajemi, ``Cost-effective maintenance tools for proprietary
  languages,'' in \emph{Proceedings IEEE International Conference on Software
  Maintenance. ICSM 2001}, 2001, pp. 240--249.

\bibitem{Deurs98a}
A.~van Deursen and P.~Klint, ``Little languages: little maintenance?'' \emph{J.
  Softw. Maintenance Res. Pract.}, vol.~10, pp. 75--92, 1998.

\bibitem{SNEED2019162}
\BIBentryALTinterwordspacing
H.~Sneed and C.~Verhoef, ``Re-implementing a legacy system,'' \emph{Journal of
  Systems and Software}, vol. 155, pp. 162--184, 2019. [Online]. Available:
  \url{https://www.sciencedirect.com/science/article/pii/S0164121219301050}
\BIBentrySTDinterwordspacing

\bibitem{Lamm01a}
R.~Lammel and C.~Verhoef, ``Cracking the 500-language problem,'' \emph{IEEE
  Software}, vol.~18, no.~6, pp. 78--88, 2001.

\bibitem{Moon01a}
L.~Moonen, ``Generating robust parsers using island grammars,'' in
  \emph{Proceedings Eight Working Conference on Reverse Engineering ({WCRE}
  2001)}, E.~Burd, P.~Aiken, and R.~Koschke, Eds.\hskip 1em plus 0.5em minus
  0.4em\relax IEEE Computer Society, Oct. 2001, pp. 13--22.

\bibitem{Moon02a}
------, ``Lightweight impact analysis using island grammars,'' in
  \emph{Proceedings 10th International Workshop on Program Comprehension},
  2002, pp. 219--228.

\bibitem{Bysi16a}
M.~Bysiek, A.~Drozd, and S.~Matsuoka, ``Migrating legacy fortran to python
  while retaining fortran-level performance through transpilation and type
  hints,'' in \emph{Proceedings of the 6th Workshop on Python for
  High-Performance and Scientific Computing}, ser. PyHPC '16.\hskip 1em plus
  0.5em minus 0.4em\relax IEEE Press, 2016, p. 9–18.

\bibitem{Fox97a}
\BIBentryALTinterwordspacing
G.~Fox, X.~Li, Z.~Qiang, and W.~Zhigang, ``\BIBforeignlanguage{en}{A prototype
  of {Fortran}-to-{Java} converter},''
  \emph{\BIBforeignlanguage{en}{Concurrency: Practice and Experience}}, vol.~9,
  no.~11, pp. 1047--1061, Nov. 1997. [Online]. Available:
  \url{https://onlinelibrary.wiley.com/doi/abs/10.1002/%28SICI%291096-9128%28199711%299%3A11%3C1047%3A%3AAID-CPE348%3E3.0.CO%3B2-V}
\BIBentrySTDinterwordspacing

\bibitem{Seym01a}
\BIBentryALTinterwordspacing
K.~Seymour and J.~Dongarra, ``Automatic translation of fortran to jvm
  bytecode,'' in \emph{Proceedings of the 2001 Joint ACM-ISCOPE Conference on
  Java Grande}, ser. JGI '01.\hskip 1em plus 0.5em minus 0.4em\relax New York,
  NY, USA: Association for Computing Machinery, 2001, p. 126–133. [Online].
  Available: \url{https://doi.org/10.1145/376656.376833}
\BIBentrySTDinterwordspacing

\bibitem{Dool99a}
D.~Doolin, J.~Dongarra, and K.~Seymour, ``\BIBforeignlanguage{English}{Jlapack
  - compiling lapack fortran to java},''
  \emph{\BIBforeignlanguage{English}{Scientific Programming}}, vol.~7, no.~2,
  pp. 111--138, 1999.

\bibitem{Feld95a}
\BIBentryALTinterwordspacing
S.~I. Feldman, ``A fortran to c converter,'' \emph{SIGPLAN Fortran Forum},
  vol.~9, no.~2, p. 21–22, oct 1990. [Online]. Available:
  \url{https://doi.org/10.1145/101363.101366}
\BIBentrySTDinterwordspacing

\bibitem{Feldm91}
\BIBentryALTinterwordspacing
S.~I. Feldman, D.~M. Gay, M.~W. Maimone, and N.~L. Schryer, ``Availability of
  {F}2c-a {Fortran} to {C} {Converter},'' \emph{SIGPLAN Fortran Forum},
  vol.~10, no.~2, pp. 14--15, Jul. 1991. [Online]. Available:
  \url{http://doi.acm.org/10.1145/122006.122007}
\BIBentrySTDinterwordspacing

\bibitem{Gros12a}
\BIBentryALTinterwordspacing
R.~W. Grosse-Kunstleve, T.~C. Terwilliger, N.~K. Sauter, and P.~D. Adams,
  ``Automatic fortran to c++ conversion with fable,'' \emph{Source Code for
  Biology and Medicine}, vol.~7, no.~1, p.~5, may 2012. [Online]. Available:
  \url{https://doi.org/10.1186/1751-0473-7-5}
\BIBentrySTDinterwordspacing

\bibitem{f2cpp}
``{F2CPP: a Python script to convert Fortran 77 to C++ code},''
  \url{http://sourceforge.net/projects/f2cpp/}, accessed: June 16, 2023.

\bibitem{Frea81a}
\BIBentryALTinterwordspacing
R.~A. Freak, ``A fortran to pascal translator,'' \emph{Softw. Pract. Exp.},
  vol.~11, no.~7, pp. 717--716, 1981. [Online]. Available:
  \url{https://doi.org/10.1002/spe.4380110708}
\BIBentrySTDinterwordspacing

\bibitem{Cari92a}
\BIBentryALTinterwordspacing
R.~B. Caringal and P.~M. Dung, ``A fortran iv to quickbasic translator,''
  \emph{SIGPLAN Not.}, vol.~27, no.~2, p. 75–87, feb 1992. [Online].
  Available: \url{https://doi.org/10.1145/130973.130979}
\BIBentrySTDinterwordspacing

\bibitem{Slap83a}
\BIBentryALTinterwordspacing
J.~K. Slape and P.~J.~L. Wallis, ``Conversion of fortran to ada using an
  intermediate tree representation,'' \emph{Comput. J.}, vol.~26, no.~4, p.
  344–353, nov 1983. [Online]. Available:
  \url{https://doi.org/10.1093/comjnl/26.4.344}
\BIBentrySTDinterwordspacing

\bibitem{Pars88a}
\BIBentryALTinterwordspacing
M.~Parsian, B.~Basdell, Y.~Bhayat, I.~Caldwell, N.~Garland, B.~Jubanowsky, and
  J.~Robinette, ``Ada translation tools development: Automatic translation of
  fortran to ada,'' \emph{Ada Lett.}, vol. VIII, no.~6, p. 57–71, nov 1988.
  [Online]. Available: \url{https://doi.org/10.1145/51634.51637}
\BIBentrySTDinterwordspacing

\bibitem{Prud77a}
\BIBentryALTinterwordspacing
J.~A. Prudom and M.~A. Hennell, ``Some problems concerning the automatic
  translation of fortran to algol 68,'' in \emph{Proceedings of the Strathclyde
  {ALGOL} 68 Conference, Glasgow, Scotland, March 29-31, 1977}.\hskip 1em plus
  0.5em minus 0.4em\relax {ACM}, 1977, pp. 138--143. [Online]. Available:
  \url{https://doi.org/10.1145/800238.807153}
\BIBentrySTDinterwordspacing

\bibitem{Feld79a}
J.~Feldman, ``High level programming for distributed computing,'' \emph{CACM},
  vol.~22, no.~6, Jun. 1979.

\bibitem{Zhan97a}
\BIBentryALTinterwordspacing
G.~Zhang, B.~Carpenter, G.~C. Fox, X.~Li, X.~Li, and Y.~Wen, ``Pcrc-based {HPF}
  compilation,'' in \emph{Languages and Compilers for Parallel Computing, 10th
  International Workshop, LCPC'97, Minneapolis, Minnesota, USA, August 7-9,
  1997, Proceedings}, ser. Lecture Notes in Computer Science, Z.~Li, P.~Yew,
  S.~Chatterjee, C.~Huang, P.~Sadayappan, and D.~C. Sehr, Eds., vol.
  1366.\hskip 1em plus 0.5em minus 0.4em\relax Springer, 1997, pp. 204--217.
  [Online]. Available: \url{https://doi.org/10.1007/BFb0032693}
\BIBentrySTDinterwordspacing

\bibitem{zhen94a}
Q.~Zheng, ``\BIBforeignlanguage{Chinese}{A fortran to c translator based on
  sigma system},'' PACT, Technical Report, 1994.

\bibitem{anqu20a}
N.~Anquetil, A.~Etien, M.~H. Houekpetodji, B.~Verhaeghe, S.~Ducasse,
  C.~Toullec, F.~Djareddir, J.~Sudich, and M.~Derras, ``Modular moose: A new
  generation of software reengineering platform,'' in \emph{International
  Conference on Software and Systems Reuse (ICSR'20)}, ser. LNCS, no. 12541,
  Dec. 2020.

\bibitem{Brag23a}
\BIBentryALTinterwordspacing
S.~Bragagnolo, ``{A Holistic Approach to Migrate Industrial Legacy Systems},''
  Theses, {Universite de Lille ; Inria}, May 2023. [Online]. Available:
  \url{https://inria.hal.science/tel-04132315}
\BIBentrySTDinterwordspacing

\bibitem{Verh19a}
B.~Verhaeghe, A.~Etien, N.~Anquetil, A.~Seriai, L.~Deruelle, S.~Ducasse, and
  M.~Derras, ``{GUI} migration using {MDE} from {GWT} to {Angular} 6: An
  industrial case,'' in \emph{2019 IEEE 26th International Conference on
  Software Analysis, Evolution and Reengineering (SANER'19)}, Hangzhou, China,
  2019, pp. 579--583.

\bibitem{Orch13a}
\BIBentryALTinterwordspacing
D.~A. Orchard and A.~C. Rice, ``Upgrading fortran source code using automatic
  refactoring,'' in \emph{Proceedings of the 2013 {ACM} Workshop on Refactoring
  Tools, WRT@SPLASH 2013, Indianapolis, IN, USA, October 27, 2013}, E.~R.
  Murphy{-}Hill and M.~Sch{\"{a}}fer, Eds.\hskip 1em plus 0.5em minus
  0.4em\relax {ACM}, 2013, pp. 29--32. [Online]. Available:
  \url{https://doi.org/10.1145/2541348.2541356}
\BIBentrySTDinterwordspacing

\bibitem{Parr07a}
T.~Parr, \emph{The Definitive {ANTLR} Reference: Building Domain-Specific
  Languages}.\hskip 1em plus 0.5em minus 0.4em\relax Pragmatic Programmers, May
  2007.

\bibitem{Quin11a}
D.~Quinlan and C.~Liao, ``The {ROSE} source-to-source compiler
  infrastructure,'' in \emph{Cetus users and compiler infrastructure workshop,
  in conjunction with PACT}, vol. 2011.\hskip 1em plus 0.5em minus 0.4em\relax
  Citeseer, 2011, p.~1.

\bibitem{Kurs13a}
J.~Kur\v{s}, G.~Larcheveque, L.~Renggli, A.~Bergel, D.~Cassou, S.~Ducasse, and
  J.~Laval, ``{PetitParser}: Building modular parsers,'' in \emph{Deep Into
  Pharo}.\hskip 1em plus 0.5em minus 0.4em\relax Square Bracket Associates,
  Sep. 2013, p.~36.

\end{thebibliography}

\end{document}